\documentclass[%
superscriptaddress,
preprint,
 amsmath,amssymb,
 prl
]{revtex4-2}

\usepackage{xcolor}
\usepackage{graphicx}% Include figure files
\usepackage{dcolumn}% Align table columns on decimal point
\usepackage{bm}% bold math
\usepackage{setspace}
\begin{document}

%\preprint{APS/PRL}

\title{White Paper: AWAKE, Plasma Wakefield Acceleration of Electron Bunches for Near and Long Term Particle Physics Applications}% Force line breaks with \\

\author{P.~Muggli}
\email{muggli@mpp.mpg.de}
\affiliation{Max Planck Institute for Physics, Munich, Germany}
\collaboration{AWAKE Collaboration}

\date{\today}% It is always \today, today,
%              %  but any date may be explicitly specified

%\begin{abstract}
%%%%
%\end{abstract}

\maketitle

%\tableofcontents

%\newpage

% something about also using X-band
% something about international collaboration ..

%%%%%%%%%%%%%%%%%%%%%%%%%%%%%%%%%%%%%%%%%%%%%%%%%%%%%%%%%%%%%%%%%%%%%%%%%%%%%%%%%%%%%%%%%%%%%%%%%%%%%%%%%%%%%%%%%%%%%%%%%%%%

\section{Executive summary}
Plasma-based accelerators have made remarkable progress over the last two decades. %
Their unique characteristics make them tools that can revolutionize fields of science and applications. %
AWAKE takes advantage of the availability of high-energy, relativistic proton bunches to drive large amplitude wakefields ($\sim$GV/m) in a single plasma over distances sufficient to produce hundreds of GeV to TeV electron bunches. % over only a few hundreds of meters or over a few kilometers, respectively. %
Hundreds of GeV bunches with $\mathcal{O}{(10^9)}$ electrons could replace current electron sources based on the generation of secondary particles, thereby extending in the next decade the reach of dark photon search experiments. %
Collisions between $\mathcal{O}{(50\,GeV)}$ electrons and protons with LHC energy would allow studying \emph{low x} processes, processes with high cross-section, making studies interesting even with low luminosity. %

Bunches of TeV electrons with TeV protons would allow for center-of-mass energy of 9\,TeV, 30 times higher than at HERA. %
Such a very-high energy ep/eA collider, would be smaller and give access to physics beyond the reach of colliders using other electron acceleration schemes. % 

Based on experimental and numerical simulation results obtained in the early phase of AWAKE (Run 1), that demonstrated the basic features of the accelerator scheme, we developed a clear science roadmap that, at the beginning of the next decade (end of Run 2), would put AWAKE in a position to propose first particle physics experiments with 50-200\,GeV electron bunches. %
This raodmap based on a plasma for self-modulation of the p$^+$ bunch and one for acceleration of the electron bunch, and on external injection in order to preserve the quality of the bunch that is accelerated (charge, energy spread, emittance). %
The roadmap includes: 
\begin{itemize}
\item 2021-22: Demonstrating seeding of the self-modulation (SM) of the p$^+$ bunch with an electron bunch; 
\item 2023-25: Introducing a step in the plasma density at a location during the SM growth for wakefields to maintain large amplitude over long plasma distances; 
\item 2028-30: External injection in the accelerator plasma of an electron bunch with parameters suitable full blowout of plasma electrons, loading of the wakefields, and matching to the plasma focusing force;
\item 2021-...: Development of a plasma source that can be extended to hundreds to thousands of meters in length and with sub-\%density uniformity. %
This is the main R\&D required for the application of this acceleration scheme to HEP. %
\end{itemize}

Besides following this roadmap, AWAKE also studies many aspects of plasma-based accelerators that are common to all schemes: external injection, emittance preservation, hosing instability, development of suitable diagnostics, development of simulation tools, and others. %
We plan to develop and use a 150\,MeV X-band linac as injector and explore even more advanced schemes, such as a laser wakefield accelerator (LWFA) injector. %

AWAKE is an open and international collaboration that would greatly benefit from the involvement of US groups, in particular on the topics of diagnostics for plasma waves and for accelerated electron bunches.

\newpage
%%%%%%%%%%%%%%%%%%%%%%%%%%%%%%%%%%%%%%%%%%%%%%%%%%%%%%%%%%%%%%%%%%%%%%%%%%%

%%%%%%%%%%%%%%%%%%%%%%%%%%%%%%%%%%%%%%%%%%%%%%%%%

\section{Introduction}
\begin{figure}[h!]
    \centering
    \includegraphics[scale=0.4]{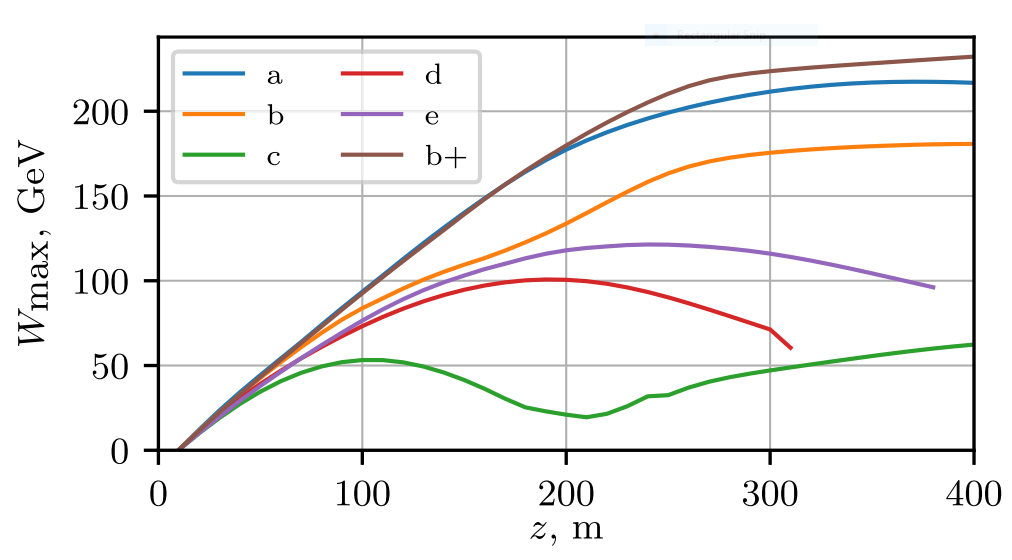}
    \includegraphics[scale=0.35]{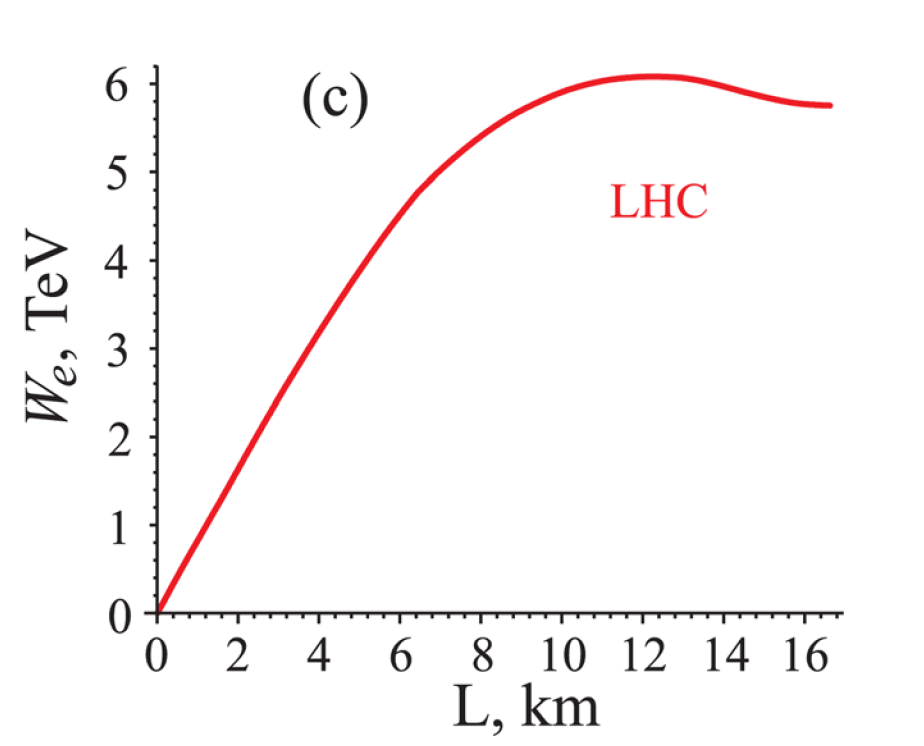}
    \caption{
    Top: the estimated maximum witness energy $W_{max}(z)$ for the different accelerator parameters, showing that an energy gain $>$200\,GeV (blue, brown lines) can be reached in $<$300\,m with a 400\,GeV p$^+$ bunch from the SPS as driver. %
    From~\cite{Lotov_2021}. %
    Bottom: the maximum energy gain of electrons $W_e$ versus propagation distance for a plasma with density step and a 7\,TeV LHC p$^+$ bunch as driver.  %
    From~\cite{bib:caldwell2011}. %
    }
    \label{fig:EgainLHC}
    \end{figure}

%PWFA: high gradient,large energy gain demonstrated
%TeV energies require scaling
%relativistic, multi-kJ p$^+$ bunches available from synchrotrons (e.g., CERN)
%large energy gain in a single plasma
Synchrotons such as the the SPS and the LHC at CERN routinely produce relativistic (400\,GeV, 7\,TeV) proton (p$^+$) bunches that also carry large amounts of energy (19, 112\,kJ). %
The availability of these drivers makes it in principle possible to drive large amplitude wakefields (GV/m) in plasma over very long distances. %~\cite{Lotov_2021}. %
Therefore an externally-injected electron witness bunch can in principle reach very large energies, hundreds of GeVs~\cite{Lotov_2021} (Fig.~\ref{fig:EgainLHC} top, 1\,GV/m over 200\,m) to TeVs~\cite{bib:caldwell2011} (Fig.~\ref{fig:EgainLHC} bottom) in a single plasma hundreds-of-meters to kilometers long. %
    %%%%%%%%%%%%%%%%%%%%%%%%%%%%%%%%%%%%%%%%%%%%%%%%%%%%%%%%%%%%%%%%%%%%%%%%%%%%%%%%%%%%%%%%%%%%%%%%%%%%%%%%%%%%%%%
%    \begin{figure}[h!]
%    \centering
%    \includegraphics[scale=0.4]{WakefieldsSPS.PNG}
%    \includegraphics[scale=0.4]{EgainSPS.PNG}
%    \caption{%The maximum longitudinal wakefield $E_{max}$ versus the propagation distance z for the selected variants, driven by an SPS p$^+$ bunch. The dotted line shows the plasma density profile $n_{e0}(z)$ for the variant ‘b+’. %
%    The estimated maximum witness energy $W_{max}(z)$ for the considered variants, showing that an energy gain $>$200\,GeV can be reached in $<$300\,m with a 400\,GeV p$^+$ bunch from the SPS as driver. %
%    From~\cite{Lotov_2021}. %
%    }
 %   \label{fig:WakefieldsSPS}
%    \end{figure}
    %%%%%%%%%%%%%%%%%%%%%%%%%%%%%%%%%%%%%%%%%%%%%%%%
    %%%%%%%%%%%%%%%%%%%%%%%%%%%%%%%%%%%%%%%%%%%%%%%%%%%%%%%%%%%%%%%%%%%%%%%%%%%%%%%%%%%%%%%%%%%%%%%%%%%%%%%%%%%%%%%
    %%%%%%%%%%%%%%%%%%%%%%%%%%%%%%%%%%%%%%%%%%%%%%%%

Based on this potential, the international AWAKE Collaboration is developing this acceleration scheme for applications to high-energy physics such as dark photon searches and QED in collisions between high-energy electrons and photons in the next decade~\cite{bib:matthew} and, in the longer term, to a very-high-energy ep/eA collider~\cite{bib:vhepp}. %

Following the scaling of plasma wakefield accelerators~\cite{bib:keinigs}, these drive p$^+$ bunches are too long ($\sim$10\,cm) to directly drive these wakefields. % 
They must self-modulate~\cite{bib:kumar} in a first plasma to be transformed into a train of bunches with period $\sim$1\,mm to resonantly drive wakefields with GV/m amplitude in a second plasma~\cite{bib:mugglirun2}. % 
Experiments with the SPS p$^+$ bunch as driver have already demonstrated the occurrence of self-modulation~\cite{bib:karl,bib:marlene}, as well as many of its characteristics~\cite{bib:fabian,bib:falk,bib:marlenesat,bib:james,bib:pablo,bib:livio}. %
They have also demonstrated that externally-injected test electrons can be accelerated (in the self-modulator) from $\sim$19\,MeV to $\sim$2\,GeV~\cite{AW:NATURE,bib:royal} over less than 10\,m~\cite{bib:marlenesat}. % 

Based on the results, a scientific roaddmap was drawn~\cite{bib:mugglirun2,bib:awakereview} to demonstrate the potential of this acceleration scheme for HEP applications by the end of the decade. %
The main ingredients of this roadmap are the use of two plasmas for SM and acceleration, and external injection of a relativistic electron bunch to demonstrate that a bunch with quality sufficient for first application can be produced: charge, energy and energy spread, and emittance. %  
The major steps in this roadmap include: the seeding of SM with an electron bunch; the use of a step in the plasma density so that wakefields maintain a large amplitude past saturation of SM; the external injection of an electron bunch with parameters suitable to reach blow-out of plasma electrons, to load the wakefields and to be matched to the plasma focusing force; and to develop plasma sources whose length can be extended and is scalable for large energy gain. %

%%%%%%%%%%%%%%%%%%%%%%%%%%%%%%%%%%%%%%%%%%%%%%%%%%%%%%%%%%%%%%%%%%%%%%%%%%%%%%%%%%%%%%%%%%%%%%%%%%%%%%%%%%%%%%%%%%%%%%%%%%%%

\section{Background}

\begin{figure}[h!]
    \centering
    \includegraphics[scale=0.33]{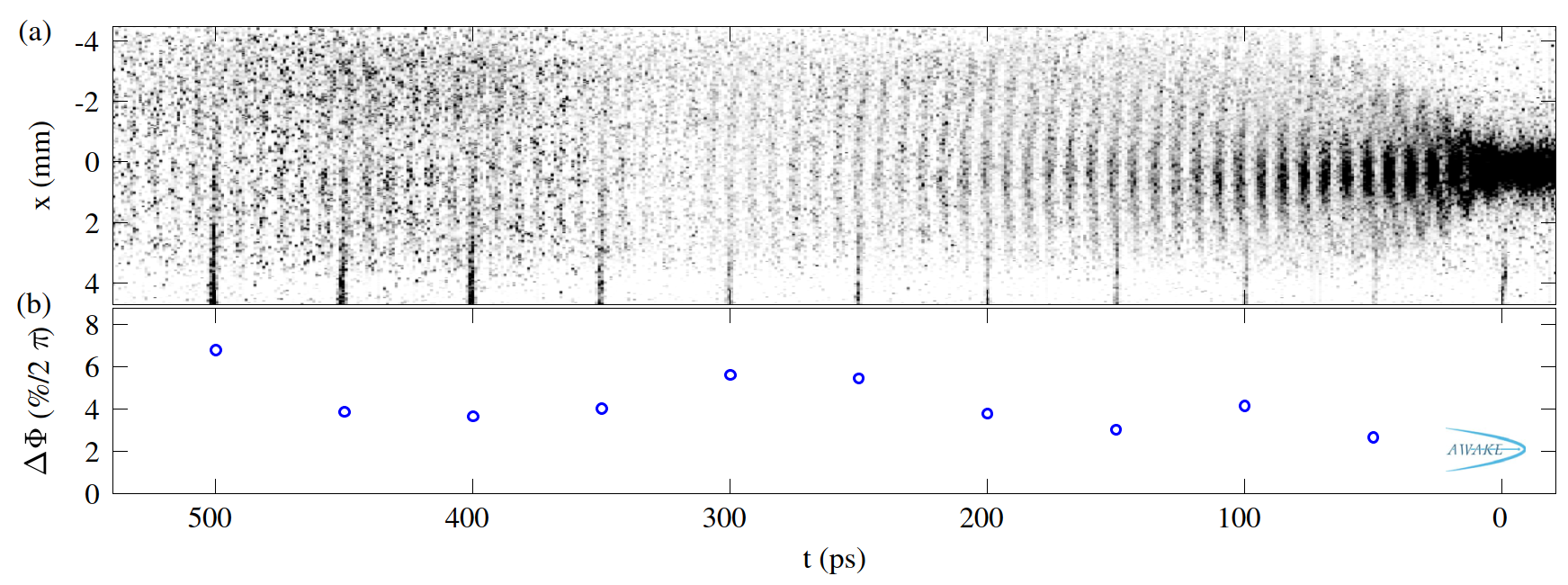}
    \caption{(a) Time-resolved, “stitched” image (11 series of $\sim$10 images or events) of the self-modulated proton bunch with the relativistic ionization front (RIF) $t_{RIF}=125\,ps$ (0.5 rms bunch length) ahead of the bunch center, $n_{e0}\cong1.81\times10^{14}$\,cm$^{-3}$. %
    The RIF is at $t=0$ (not visible), time reference signal visible every 50\,ps at the bottom of the image. %
    (b) The rms phase variation between the modulation along the bunch and the time reference signal.  %
    Variations are limited to a small relative fraction ($<$7\%) of the $2\pi$ phase of the period of the wakefields. %
    From~\cite{bib:fabian}. %
    }
    \label{fig:SSM}
    \end{figure}

The occurrence of SM was observed (fig.~\ref{fig:SSM}, top)~\cite{bib:karl,bib:marlene}. %
Modulation of the bunch density is at the period of the wakefields in the plasma of electron density $n_{e0}$: $\tau_M\cong1/f_{pe}$, $f_{pe}=\left(n_{e0}e^2/\epsilon_0 m_e\right)^{1/2}/2\pi$~\cite{bib:karl}. %
The amplitude of the wakefields grows both along the plasma and along the bunch. %
Their amplitude reaches hundreds of MV/m, much larger than their initial value ($<$10\,MV/m)~\cite{bib:marlene}. %
Most importantly, the predicted instability~\cite{bib:kumar} (or SMI) can be seeded and made reproducible and controllable using a relativistic ionization front (RIF) placed within the p$^+$ bunch (fig.~\ref{fig:SSM}, bottom)~\cite{bib:fabian}. %
Self-modulation is seeded (SSM) only when the amplitude of the wakefields exceed a threshold value of $\sim$4\,MV/m ($n_{e0}=10^{14}$\,cm$^{-3}$). %
Reproducibility and control of the instability are necessary for deterministic injection of an electron bunch into the accelerating \emph{and} focusing phase of the wakefields. %
    %%%%%%%%%%%%%%%%%%%%%%%%%%%%%%%%%%%%%%%%%%%%%%%%%%%%%%%%%%%%%%%%%%%%%%%%%%%%%%%%%%%%%%%%%%%%%%%%%%%%%%%%%%%%%%%
    %%%%%%%%%%%%%%%%%%%%%%%%%%%%%%%%%%%%%%%%%%%%%%%%

\begin{figure}[h!]
    \centering
    \includegraphics[scale=0.35]{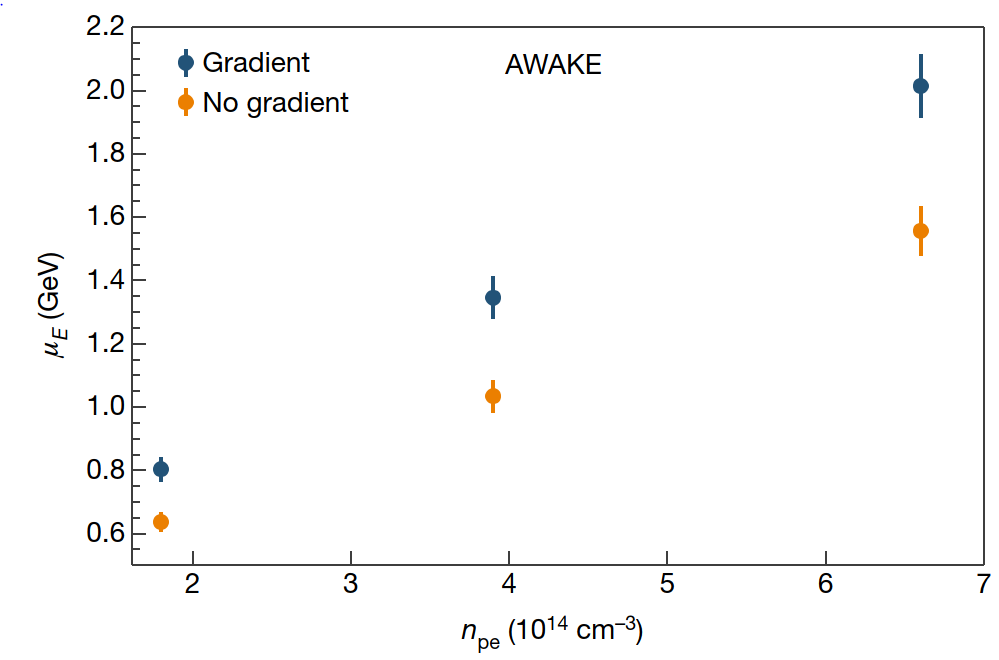}
    \caption{Measurement of the highest peak energies $\mu_E$ achieved at different plasma densities $n_{pe}$, with and without a gradient in the plasma density. %
    The error bars arise from the position–energy conversion. %
    The gradients chosen are those that were observed to maximize the energy gain. %
    Acceleration to 2.0$\pm$0.1\,GeV is achieved with a plasma density of $n_{e0}\cong6.6\times10^{14}$\,cm$^{-3}$ with a density difference of $\pm$2.2\% $\pm$0.1\% over 10\,m.  %
    From~\cite{AW:NATURE}. %
    }
    \label{fig:EgainExp}
    \end{figure}

Measurements show that the wakefields have a phase velocity smaller than the velocity of the p$^+$ bunch~\cite{bib:falk}, as expected from theory and simulations~\cite{bib:pukhov,bib:schroeder}. %
Protons defocused by the SM process provide details on the development of SM as well as on the fine scale evolution of the modulation period along the plasma~\cite{bib:pablo}. %
The SM saturates a distance between 3 and 5\,m into the plasma~\cite{bib:marlenesat}. %
Side-injected, test electrons with an energy of $\sim$19\,MeV were accelerated to $\sim$2\,GeV in less than 10\,m of plasma with density $n_{e0}\cong7\times10^{14}$\,cm$^{-3}$ (Fig.\ref{fig:EgainExp})~\cite{AW:NATURE}. %
The SM is in general well understood and there is excellent agreement between experimental and simulation results~\cite{bib:gorn,bib:pablo}. %
    %%%%%%%%%%%%%%%%%%%%%%%%%%%%%%%%%%%%%%%%%%%%%%%%%%%%%%%%%%%%%%%%%%%%%%%%%%%%%%%%%%%%%%%%%%%%%%%%%%%%%%%%%%%%%%%
    %%%%%%%%%%%%%%%%%%%%%%%%%%%%%%%%%%%%%%%%%%%%%%%%

%%%%%%%%%%%%%%%%%%%%%%%%%%%%%%%%%%%%%%%%%%%%%%%%%%%%%%%%%%%%%%%%%%%%%%%%%%%%%%%%%%%%%%%%%%%%%%%%%%%%%%%%%%%%%%%%%%%%%%%%%%%%

\section{Roadmap towards particle physics applications}

    %%%%%%%%%%%%%%%%%%%%%%%%%%%%%%%%%%%%%%%%%%%%%%%%%%%%%%%%%%%%%%%%%%%%%%%%%%%%%%%%%%%%%%%%%%%%%%%%%%%%%%%%%%%%%%%
    \begin{figure}[h!]
    \centering
    \includegraphics[scale=0.75]{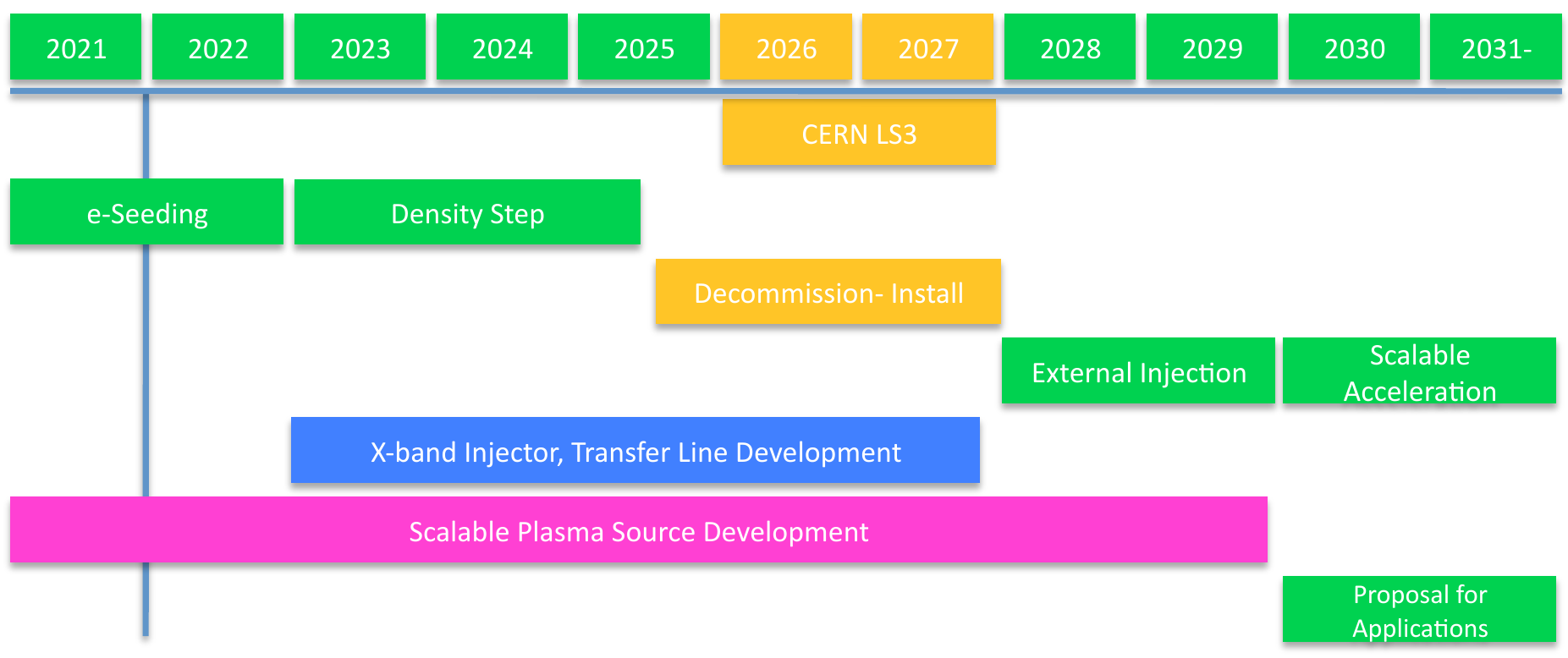}
    \caption{
    Roadmap for the AWAKE project towards application to HEP in the next decade. %
    Years in green/yellow are years with/without SPS beam available. % 
%    From~\cite{bib:caldwell2011}. %
    }
    \label{fig:roadmap}
    \end{figure}
    %%%%%%%%%%%%%%%%%%%%%%%%%%%%%%%%%%%%%%%%%%%%%%%%

\begin{figure}[h!]
    \centering
    \includegraphics[scale=0.55]{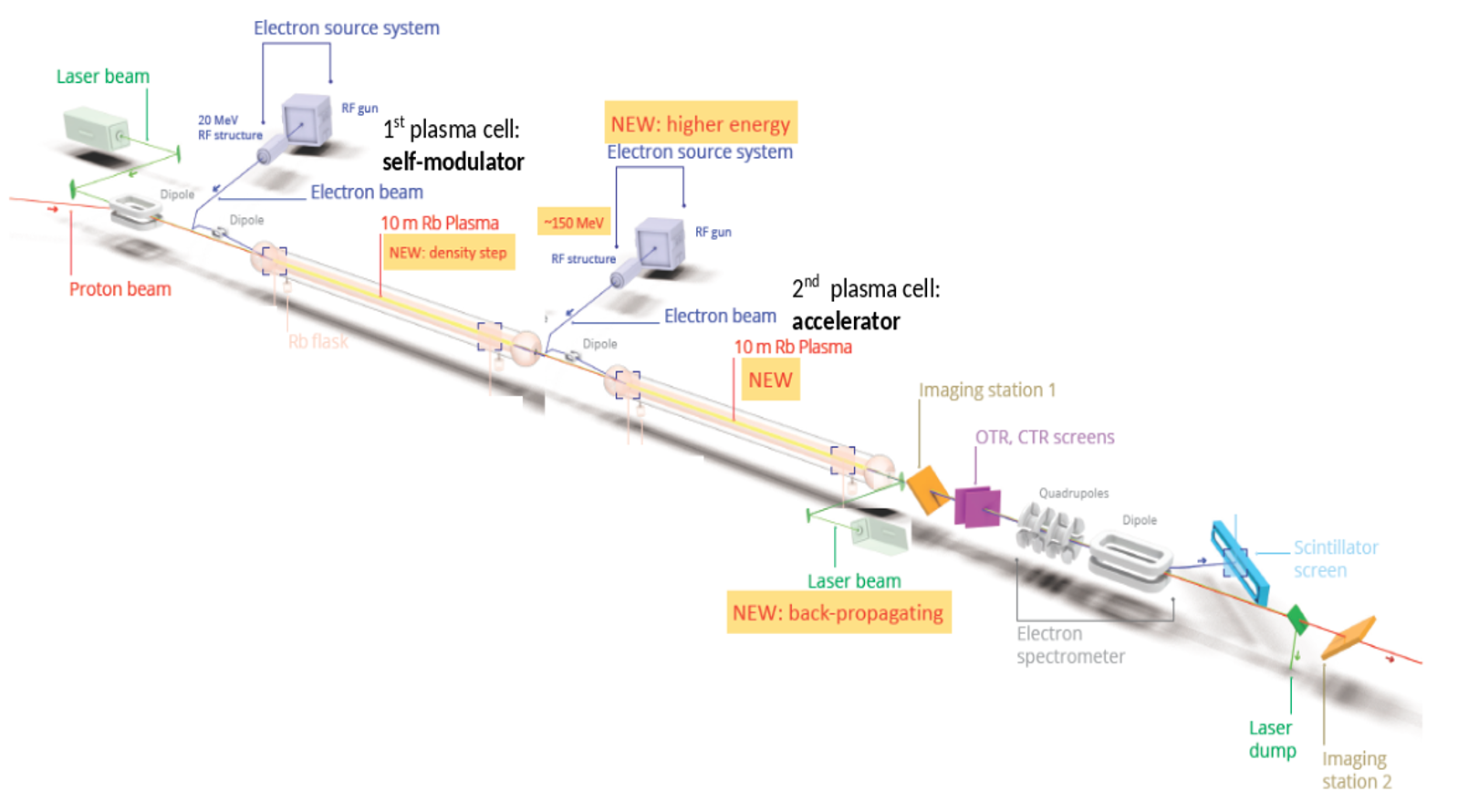}
    \caption{
    Schematic of the experimental setup for the AWAKE program. %
    The first milestones focus on the self-modulator (first) plasma. %
    The last two milestones include two plasmas $\sim$10\,m-long each. %
    Applications would use a similar set-up, with a much longer second, accelerator plasma. %
%    From~\cite{bib:caldwell2011}. %
    }
    \label{fig:run2c_scheme}
    \end{figure}

Based on the state-of-the-art, a clear science roadmap (Fig.~\ref{fig:roadmap}) can be drawn for experiments that should lead by the end of the decade to a demonstrator for a device with a 10\,m-long accelerator producing a $\sim$10\,GeV electron bunch with $\sim$100\,pC charge and energy spread (\%-level) and normalized emittance (1-10\,mm-mrad), sufficient for early HEP applications~\cite{bib:matthew}. %
The milestones along the roadmap are clear. %
They rely on developing a system with two plasmas: the first one for SM of the p$^+$ bunch and the second one for acceleration of the electron bunch in the wakefields driven by the self-modulated p$^+$ bunch (Fig.~\ref{fig:run2c_scheme}). %
This program was discussed in Ref.~\cite{bib:mugglirun2} and is constantly and further developed. %
The roadmap includes four main milestones. %

    %%%%%%%%%%%%%%%%%%%%%%%%%%%%%%%%%%%%%%%%%%%%%%%%%%%%%%%%%%%%%%%%%%%%%%%%%%%%%%%%%%%%%%%%%%%%%%%%%%%%%%%%%%%%%%%
    %%%%%%%%%%%%%%%%%%%%%%%%%%%%%%%%%%%%%%%%%%%%%%%%

    %%%%%%%%%%%%%%%%%%%%%%%%%%%%%%%%%%%%%%%%%%%%%%%%%%%%%%%%%%%%%%%%%%%%%%%%%%%%%%%%%%%%%%%%%%%%%%%%%%%%%%%%%%%%%%%%%%%%%%%%%%%%

    \subsection{1. Demonstration of SM seeding with an electron bunch}
    
    Seeding of SM with a RIF~\cite{bib:fabian} leaves the front of the p$^+$ bunch not modulated. %
    However, a very long accelerator plasma is likely to be pre-ionized. %
    This could result in the SMI of the front of the bunch in the plasma of the accelerator section. %
    Wakefields driven by this front SMI could interfere with the modulated bunch train emerging from the self-modulator. %
    In contrast, an electron bunch placed ahead of the p$^+$ bunch drive wakefields that can seed SM (eSSM)of the entire p$^+$ bunch. %
    Preliminary results show that e-SSM is possible~\cite{bib:livio}. %
    Studies of e-SSM started in 2021 and will continue in 2022. %

    %%%%%%%%%%%%%%%%%%%%%%%%%%%%%%%%%%%%%%%%%%%%%%%%%%%%%%%%%%%%%%%%%%%%%%%%%%%%%%%%%%%%%%%%%%%%%%%%%%%%%%%%%%%%%%%%%%%%%%%%%%%%

    \subsection{2. Plasma density step}
    
\begin{figure}[h!]
    \centering
    \includegraphics[scale=0.25]{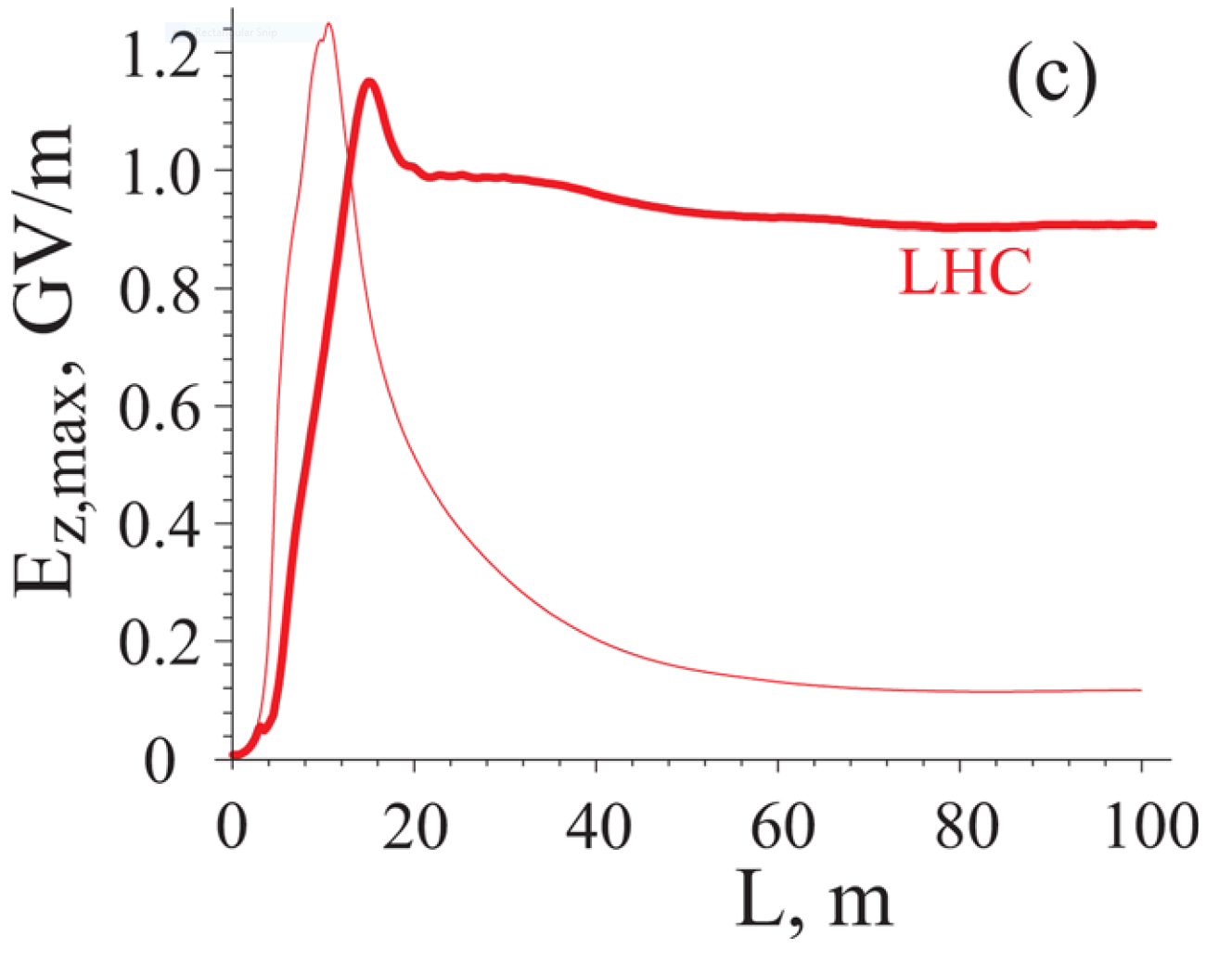}
    \caption{The maximum longitudinal wakefield $E_{z,max}$ versus propagation distance z, with an LHC p$^+$ bunch as driver, without (thin red line) and with (thick red line) a plasma density step.  %
    The density step freezes the amplitude of the wakefields close to their value near saturation ($\sim$1\,GeV/m). %
    From~\cite{bib:caldwell2011}. %
    }
    \label{fig:WakefieldsLHC}
    \end{figure}

    Numerical simulation results~\cite{bib:kumar,bib:caldwell2011} show that with a constant plasma density along the self-modulator the amplitude of the wakefields decreases after saturation of SM, because of the continuous evolution of the beam/plasma interaction. %
    They also show that introducing a plasma density step at a location along the growth of the SM allows for the wakefields to maintain an amplitude close to their saturation value over a long distance~\cite{bib:caldwell2011,bib:lotovstep} (Fig.~\ref{fig:WakefieldsLHC}). %
    The plasma source, based on ionization of a rubidium vapor by a short laser pulse (RIF)~\cite{bib:erdem,bib:erdem2} allows a temperature and thus density step to be imposed. %
    We are designing a source with variable step position and height for experiments in 2023-25 aimed at demonstrating the effect of the density on the SM and on its seeding. %
    %%%%%%%%%%%%%%%%%%%%%%%%%%%%%%%%%%%%%%%%%%%%%%%%%%%%%%%%%%%%%%%%%%%%%%%%%%%%%%%%%%%%%%%%%%%%%%%%%%%%%%%%%%%%%%%
    %%%%%%%%%%%%%%%%%%%%%%%%%%%%%%%%%%%%%%%%%%%%%%%%

    %%%%%%%%%%%%%%%%%%%%%%%%%%%%%%%%%%%%%%%%%%%%%%%%%%%%%%%%%%%%%%%%%%%%%%%%%%%%%%%%%%%%%%%%%%%%%%%%%%%%%%%%%%%%%%%%%%%%%%%%%%%%

    \subsection{3. External injection of an electron bunch}
    
    Acceleration preserving bunch quality requires injection of a low energy ($\sim$150\,MeV) electron bunch on the axis of the accelerator plasma driven by the self-modulated p$^+$ bunch emerging from the self-modulator plasma. %
    The length of the gap between the two plasmas must be minimized to decrease the evolution of the self-modulated p$^+$ bunch along the gap. %
    This evolution leads to wakefields with lower amplitude in the accelerator plasma than at the exit of the self-modulator plasma. %
    The parameters of the injected electron bunch must be suitable to drive wakefields in the blowout regime and to load the wakefields in order to preserve the emittance of the bunch and maintain a low energy spread~\cite{bib:veronica}. %
    The bunch must be matched to the focusing force of the plasma ion column to avoid deleterious effects caused by betatron oscillations of the bunch envelope. % 
    Both the in-coming (matched) beam beta-function ($\beta_0\cong$5\,mm, 150\,MeV) and out-coming one ($\beta_0\cong$40\,cm, 10\,GeV) are much shorter than the plasma length ($\ge$10\,m). %
    If not matched, variation of the phase advance of the beam over the plasma length due to small beam and plasma parameter variations from event to event would lead to mismatch of the beam to the following (magnetic) optics. %
    We note here that even though the accelerator operates at relatively low plasma density (typically $n_{e0}=7\times 10^{14}$\,cm$^{-3}$), beam parameters and transverse alignment tolerance are rather challenging to meet~\cite{bib:johnalign,bib:rebecca}. %
    
    The injector of the current design consists of an original S-band photo-injector gun, X-band linac and transfer line combination to produce $\sim$100\,pC, $\sim$200\,fs, 150\,MeV electron bunches with a beta-function of $\sim$5\,mm at the plasma entrance. %
    However, we are also investigating other accelerator schemes, such as a LWFA injector, or a foil injector. %

    %%%%%%%%%%%%%%%%%%%%%%%%%%%%%%%%%%%%%%%%%%%%%%%%%%%%%%%%%%%%%%%%%%%%%%%%%%%%%%%%%%%%%%%%%%%%%%%%%%%%%%%%%%%%%%%%%%%%%%%%%%%%

    \subsection{4. Development of long accelerator plasma source}
    
    Plasma creation using laser ionization limits the plasma length to a few tens of meters due to depletion of the laser pulse energy and to other optical constraints. %
    We are therefore investigating helicon and discharge sources for the accelerator plasma. %
    The main challenge for the accelerator plasma is the density uniformity of $<$0.5\% imposed by the resonant driving of the wakefields. %
    
\begin{figure}[h!]
    \centering
    \includegraphics[scale=0.4]{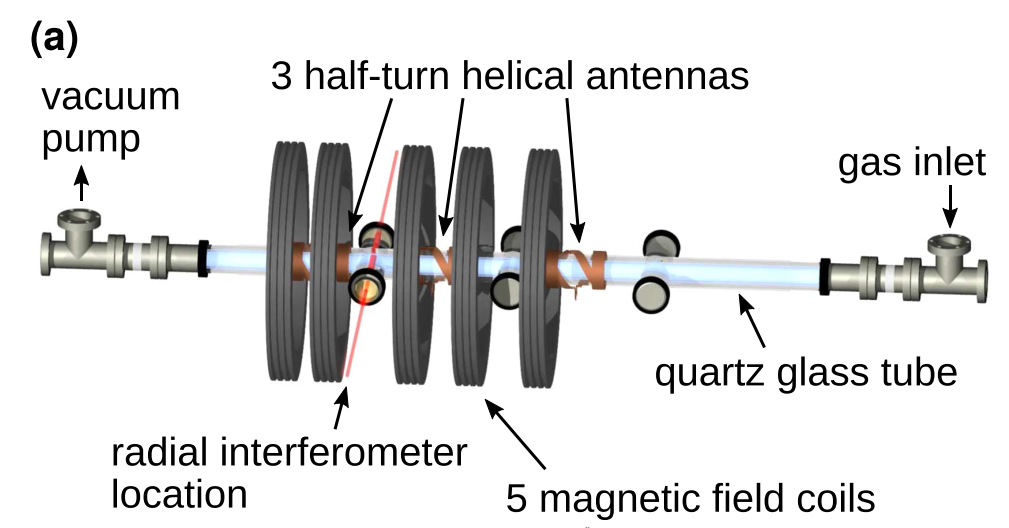}
    \caption{Schematic of the 1\,m long prototype module PROMETHEUS-A for the plasma wakeﬁeld accelerator experiment AWAKE. The magnetic ﬁeld coils are adjusted to produce a ﬁeld as homogeneous as possible, while providing access to the radial ports of the tube for diagnostic purposes. %
    From~\cite{bib:helicon}. %
    }
    \label{fig:helicon}
    \end{figure}

    The helicon source is a modular, weakly magnetized RF-discharge system whose unit cell consists of an RF antenna and of two magnetic coils (Fig.~\ref{fig:helicon}). %
    Stacking of unit cells can therefore in principle produce very long plasmas. %
    Initial experiments have shown that densities required for AWAKE can be produced~\cite{bib:helicon}. %
    %%%%%%%%%%%%%%%%%%%%%%%%%%%%%%%%%%%%%%%%%%%%%%%%%%%%%%%%%%%%%%%%%%%%%%%%%%%%%%%%%%%%%%%%%%%%%%%%%%%%%%%%%%%%%%%
    %%%%%%%%%%%%%%%%%%%%%%%%%%%%%%%%%%%%%%%%%%%%%%%%
   
    The direct-current discharge plasma source (DPS) is a simpler source. %
    However it requires stacking of multiple discharges each with start/end electrodes to reach long distances. %
    Demonstration SM experiments with a 10\,m-long DPS are tentatively planed for early 2023. % 
    
    We are developing diagnostics to measure the plasma density uniformity and reproducibility of both sources. %
    If ready, one of these sources could become the source for the accelerator plasma for the first external injection experiments, currently planned with a laser-ionized rubidium vapor source~\cite{bib:mugglirun2,bib:awakereview}. %
    
    Early application of the acceleration scheme~\cite{bib:matthew} may require an accelerator plasma tens to hundreds of meters long~\cite{Lotov_2021}, whereas eA collider applications~\cite{bib:vhepp} will require kilometers-long plasmas. %
    
    The development of long, scalable plasma sources is the main R\&D topic for the application of the acceleration scheme. %

%%%%%%%%%%%%%%%%%%%%%%%%%%%%%%%%%%%%%%%%%%%%%%%%%%%%%%%%%%%%%%%%%%%%%%%%%%%%%%%%%%%%%%%%%%%%%%%%%%%%%%%%%%%%%%%%%%%%%%%%%%%%

\section{Physics cases, applications}

Applications for high-energy electron bunches that can be produced by the AWAKE acceleration scheme are described in Refs~\cite{bib:matthew,bib:vhepp,bib:awakereview}. %
They all use a layout similar of that of Fig.~\ref{fig:run2c_scheme}. %

These applications use 50-200\,GeV electron that can be produced with p$^+$ bunches from the SPS (400\,GeV) as driver and electrons at the TeV energy scale that can be produced with the TeV p$^+$ bunch of the LHC as driver of wakefields. %
In general, the p$^+$ bunch can be used only once to drive wakefields, which limits the repetition rate of the electron  acceleration process to that of the p$^+$ bunch production. %
This leads to a significantly lower luminosity than with other electron production schemes. %
However, high-gradient acceleration permit a shorter accelerator length and the scheme can produce high-energy electrons with available or spent p$^+$ beams, offering unique opportunities in particle physics. %

\begin{figure}[bthp!]
\centering
\includegraphics[trim={5cm 0cm 6cm 1.5cm},clip,width=0.6\textwidth]{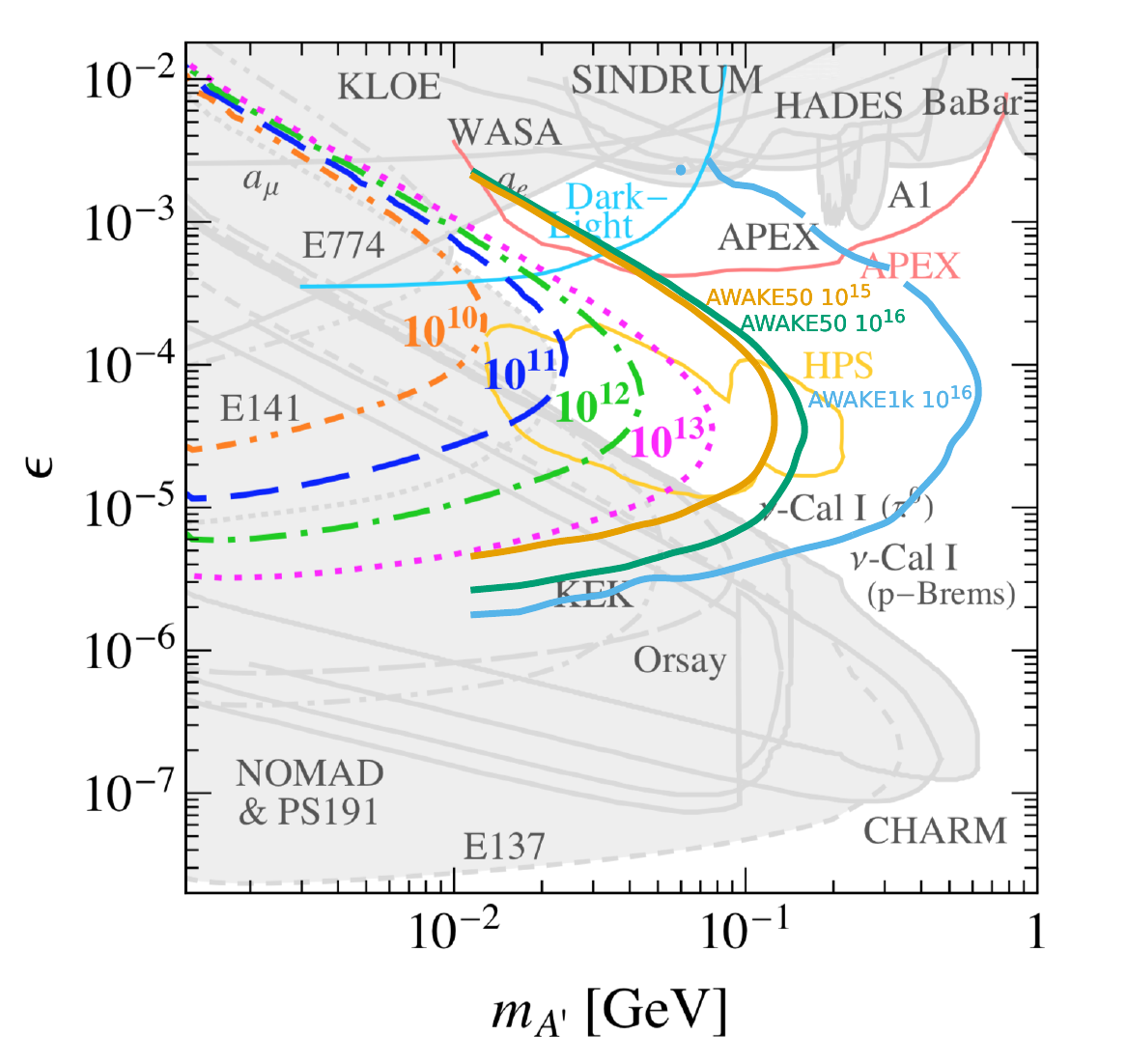}
\caption{Limits on dark photon production decaying to an $e^+e^-$ pair in terms of the mixing strength, $\epsilon$, and dark photon mass, $m_{A^\prime}$, from previous measurements (light grey shading). %
The expected sensitivity for the NA64 experiment is shown for a range of electrons on target, $10^{10} - 10^{13}$. %
Expectations from other potential experiments are shown as colored lines. %
Expected limits are also shown for $10^{15}$ (orange line) or $10^{16}$ (green line) electrons of 50\,GeV (``AWAKE50'') on target and $10^{16}$ (blue line) electrons of 1\,TeV (``AWAKE1k'') on target provided to an experiment using the future AWAKE accelerator scheme.  From Ref.~\cite{Alemany:2019vsk}.
}
\label{fig:dark_photons}
\end{figure}

Using the SPS p$^+$ bunch as driver electrons with energies up to $\sim$200\,GeV can be produced. %
These find applications in dark photon searches on fixed target in the beam dump mode ($\sim$10$^9$ electrons per bunch). %
The larger number of electrons produced at 50\,GeV than with current secondary production processes would allow for larger reach in coupling strength and dark photon mass than other other existing or planned experiments (Fig.~\ref{fig:dark_photons}). %

High-energy electron bunches could be used to investigate non-linear QED in electron-photon collisions. %
These experiments usually have a repetition rate limited by that of the high-power laser (typically 1-10\,Hz). %
The low repetition rate of the p$^+$ bunch and thus of the acceleration process may not be a limitation. %

Much higher energy, TeV-range electrons produced with an LHC p$^+$ bunch could be used for lower luminosity measurements in electron-proton or electron-ion collisions. %

%%%%%%%%%%%%%%%%%%%%%%%%%%%%%%%%%%%%%%%%%%%%%%%%%%%%%%%%%%%%%%%%%%%%%%%%%%%%%%%%%%%%%%%%%%%%%%%%%%%%%%%%%%%%%%%%%%%%%%%%%%%%

\section{Synergies with other concepts}

The p$^+$-driven plasma wakefield accelerator addresses may issues common to all plasma-based concepts. %
This includes the challenges linked to external injection, which will have to be met at the entrance of every stage of a plasma-based accelerator concept. %
Reaching transverse and time alignment tolerances requires extremely reproducible beam optics. %
Misalignment can degrade acceleration performances, but can also induce hose instability. %
Therefore studying the hose instability, understanding tolerances for his development is thus also essential for beam quality preservation in all accelerator schemes. %

Development of plasma sources with sufficient reproducibility and characteristic is essential for all plasma-based accelerator concepts. %

The interaction with, and the collaboration of US partners and institutions would be a great benefit. %
We note here that the p$^+$-driven concept could also be developed with beams available at the Brookhaven National Laboratory and also find applications there. %
Moreover, the near-term applications proposed here can be essential to increase the credibility of all plasma-based accelerator schemes for application to particle physics. %

%%%%%%%%%%%%%%%%%%%%%%%%%%%%%%%%%%%%%%%%%%%%%%%%%%%%%%%%%%%%%%%%%%%%%%%%%%%%%%%%%%%%%%%%%%%%%%%%%%%%%%%%%%%%%%%%%%%%%%%%%%%%

\section{Conclusions}

Proton-driven plasma wakefield acceleration has the potential to produce very high energy electron bunches for a number of HEP applications. %
Proton bunches suitable to drive the accelerator are available today. %
The concept is developed in the context of an international collaboration that would benefit from US contributions. %
The roadmap and mid- and long-term applications to HEP described in this white paper are based on an active experimental, simulation and theory program. % with participation of a large international collaboration. %
The main R\&D required for applications is on plasma sources scalable to very long lengths (hundreds thousands of meters) with sub-\% level plasma density uniformity. %
The program addresses many issues common to all plasma-based acceleration concepts. %
Applications to collisions with fixed targets and p$^+$ bunches relax the electron bunch parameters (in particular emittance) and operation even with luminosity lower than that required for example for an e$^-$e$^+$ is attractive. %
Therefore, first applications to HEP could be become possible early in the next decade. %

%%%%%%%%%%%%%%%%%%%%%%%%%%%%%%%%%%%%%%%%%%%%%%%%%%%%%%%%%%%%%%%%%%%%%%%%%%%%%%%%%%%%%%%%%%%%%%%%%%%%%%%%%%%%%%%%%%%%%%%%%%%%
\bibliography{biblio}% Produces the bibliography via BibTeX.

\end{document}